\documentstyle[12pt,aasms4]{article} \voffset = -8mm 
\def\etal{et al. }
\begin{document}

\title{Spectra of Southern Pulsars}

\author{ M. Toscano$^{1,2}$, M. Bailes$^{2}$, R. N. Manchester$^3$, J. S. Sandhu$^4$}
\bigskip
\bigskip
\noindent
\begin{footnotesize}
\noindent
$^1$Physics Department, University of Melbourne, Parkville, Vic 3052,
Australia.\\ $^2$Astrophysics and Supercomputing, Mail No. 31, Swinburne University of
Technology, PO Box 218, Hawthorn, Vic 3122, Australia.\\
$^3$Australia Telescope National Facility, CSIRO, PO Box
76, Epping, NSW 2121, Australia.\\ $^4$Department of Astronomy,
Caltech, Pasadena CA 91125. \\

\end{footnotesize}
\bigskip
\bigskip

\abstract{ We compare the spectral properties of the millisecond and
slow pulsars detected in the Parkes 70 cm survey. The mean spectral
index for the  millisecond pulsars (MSPs) is --1.9$\pm$0.1 whereas the
mean spectral  index for the slow pulsars is a surprisingly steep
--1.72$\pm$0.04.  A Kolmogorov-Smirnov test indicates that there is
only a 72\% probability  that the two distributions differ. As a
class, MSPs are therefore only  fractionally steeper-spectrum objects
than slow pulsars, as recent literature would suggest. We then model 
the expected distribution of millisecond pulsars in the Galaxy and find that
high-frequency surveys, with  sensitivities similar to the current
Parkes multibeam survey, are likely to detect MSPs in large
numbers. The observed distribution of MSPs will be  much less
isotropic than that resulting from low-frequency surveys,  with 50\%
of detectable MSPs residing within 11\arcdeg of the Galactic  plane in
an all-sky survey.}


\newpage 

\section{Introduction}  

In the early 1980s, attention was focussed on the enigmatic radio
complex at the position of 4C21.53. In particular the object to the
west of this position, named 4C21.53W, was resolved into two
components,  an extended flat-spectrum source $\sim$1 arcsec north of
a compact steep-spectrum object. The steep spectrum and interplanetary
scintillation of the compact source suggested that it was a radio
pulsar. Several searches, sensitive only to  periods greater than a
few milliseconds, failed to detect any pulsations,  but Backer \etal
(1982) \nocite{bkh+82} announced that the source was the first
millisecond pulsar (MSP), PSR B1937+21, with a rotation period of just
1.5 ms. Spurred on by this exciting discovery, Hamilton, Helfand \&
Becker (1985) \nocite{hhb85} made a spectral survey in 12 nearby
globular clusters for unresolved objects which might be MSPs. Their
best candidate, in the core of M28, was shown to be a highly linearly
polarized, steep-spectrum object (\cite{embh87}). This steep-spectrum
source was later found to be the fourth MSP (PSR B1821--24) and the
first globular cluster pulsar discovered (\cite{lbm+87}). A search
aimed at a further 24 nearby globular clusters resulted in the
discovery  of yet another steep-spectrum pulsar, PSR B1620--26, in M24
(\cite{lbb+88}).

The first detailed spectral study of MSPs was made by Foster, Fairhead
\& Backer (1991; FFB)\nocite{ffb91}.  Their study included the three
pulsars mentioned above, as well as PSR B1855+09. The latter was
discovered in a 430 MHz pulsar survey conducted at Arecibo
(\cite{srs+86}; \cite{fru89}). Except for PSR B1855+09, all the MSPs
had spectral indices steeper than --2.3. Although the sample of four
pulsars was not very statistically significant, this paper helped
reinforce the growing belief that MSPs had spectra that were
significantly steeper than their slower counterparts.

Since then there have been a number of large-scale surveys for pulsars
(e.g., \cite{bl92}; \cite{clj+92}; \cite{fcwa95}; \cite{cnt93}). As a
result there are now some 60 known pulsars with periods less than 20
ms. The high-frequency surveys of Clifton \& Lyne (1986) and Johnston
\etal (1992) \nocite{cl86,jlm+92} were mildly sensitive to  MSPs but
found none in their surveys of the Galactic plane.  This restricted
the population of very luminous high-frequency MSPs,  but offered
little insight into the number of low-luminosity MSPs  detectable at
high frequencies. On the other hand, the Parkes 70 cm survey
(\cite{mld+96}; \cite{lml+98}) detected 19 MSPs, 17 of which were new
discoveries. The history of MSP searching with its focus on
steep-spectrum objects, and the comparative success of low-frequency
compared with high-frequency surveys  might lead one to conclude that
there are good reasons to  conduct MSP searches exclusively at low
frequencies.

The major works on pulsar spectra are those of Lorimer \etal (1995)
\nocite{lylg95} and Kramer \etal (1998) \nocite{kxl+98}. Lorimer \etal
derived a mean spectral index for millisecond and slow pulsars of --2
and  --1.6 respectively. Their study included all of the pulsars
regularly  observed from Jodrell Bank, and did not differentiate
between  those  found in high- or low-frequency surveys. Kramer \etal
attempted to reduce observational biases by selected spectra only from
pulsars out to a distance of 1.5 kpc. They found the average spectra of
MSPs and slow pulsars to be essentially the same, with mean indices of
--1.6 and --1.7 respectively. As noted by Kramer \etal the  Parkes
survey, with its large number of detections of both millisecond and
slow pulsars,  provides an excellent sample from which to discuss the
spectral properties of pulsars in a more unbiased way.

All of the 19 MSPs detected in the Parkes survey have been either
timed regularly at Parkes or observed for polarization studies with
the Caltech pulsar correlator (\cite{nav94}). The correlator provides
accurate fluxes suitable for a spectral study. The aim of this paper
is to compare the spectra of the millisecond and slow pulsars detected
in the Parkes survey. In \S 2 of this paper we describe the
observations and data-reduction techniques used in obtaining the
spectra. \S 3 presents the average flux density measurements and
spectral indices for the MSPs and in \S 4 we compare  the spectral
index distribution of these MSPs and the slow pulsars  detected by the
survey. This demonstrates that MSPs found in a large-scale,
low-frequency survey have spectra which are only slightly steeper than
their slower counterparts, and supports the case for high-frequency 
surveys for MSPs. In \S 5 we present the results of a simulated high-frequency
search for MSPs near the Galactic plane which suggests that MSPs will
be found in significant numbers by the current Parkes multibeam survey.

\section{Observations and Data Reduction}

The 19 MSPs detected in the Parkes survey were observed using the 64 m
radio telescope at Parkes in 12 sessions between 1996 February and 
October. These observations form a subset of data taken in a
continuing pulsar  timing program. PSR B1620--26 is not timed regularly
at Parkes, but was observed on 10 occasions. The remaining 18 MSPs
were observed much more  frequently, and our data set consists of
approximately 1700 observations  each of 24 min duration. The large
number of observations and their spread  over an extended time span
should reduce the observational biases introduced by refractive and
diffractive scintillation.

Our observations were made using three dual-channel cryogenic receiver
systems with center frequencies of 436, 660 and 1500 MHz.  At the
lower frequencies the system bandwidth was 32 MHz, but for the
highest-frequency system, two bands of 128 MHz bandwidth centered at
1400 and 1660 MHz were used.  System equivalent flux
densities for the three receivers were approximately 100, 90 and 40 Jy
respectively.  The flux-density scale was established using
observations of Hydra A which was assumed to have a flux density of
134.60 Jy at 400 MHz and a spectral index of --0.909 (\cite{bgpw77}).
The system gain of all three receivers was monitored using a
linearly polarized pulse calibration signal, injected at $45\arcdeg$
to the two feed probes.

The data were two-bit digitized and autocorrelation functions (ACF)
computed using the Caltech correlator. These were integrated at the
apparent pulsar period in memory giving 1024 samples per pulsar period
for each of 512 lags. Data were typically integrated for 90 s in the
hardware integrator and then transferred to a Sun Sparc-20 computer
for further processing. The ACF data were corrected for
non-linearities resulting from the 2-bit digitisation, and Fourier
transformed to form power spectra for each of the 90 s integrations
and pulse phase bins; the spectra were dedispersed into 16 or 32
frequency channels, and stored on disk.

A typical observation provided us with sixteen, 90-s integrations in
each of two polarizations and several frequency sub-bands. The number
of bins across each profile was dependent upon the amount of
dispersion smearing across the band and the duty cycle of the
pulsar. To obtain flux densities, profiles were added in time and
frequency  and the two polarizations summed. After baseline removal,
the mean flux  density for that observation was computed.

For each pulsar, the unweighted mean flux density of the measurements
at a particular observing frequency was computed, as well as the
standard error, ${\sigma}_{S} = {\sigma}/{\sqrt{N-1}}$, where $\sigma$
is the standard deviation of the $N$ measurements about their mean. We
found that the spectra of the  MSPs in our sample were often well
described by a power-law distribution of the form  $S =
S_{0}{\nu}^{\alpha}$ as illustrated in the plot of typical spectra in
Figure 1.  A ${\chi}^2$ minimisation technique was used to fit for the
spectral index, $\alpha$,  for each pulsar. We calculated the standard
error, ${\sigma}_{\alpha}$, for each spectral  index from the variance
in $\alpha$ returned by the least-squares fitting program.

To make a meaningful comparison with the slow pulsars, we limited our
sample of slow pulsars to those detected in the Parkes 70 cm survey:
279 in total. In this way, we are able to compare in a relatively
unbiased manner the spectral indices of slow and millisecond pulsars.
Follow-up observations of the slow pulsars have not been as thorough
as for the MSPs, and 63 of them have flux density measurements only
at frequencies near 430 MHz. Of the remaining 216 for which we have
calculated spectral indices, 117 had measurements at three or more
frequencies.

\section{Results}

Table 1 shows the flux densities measured for southern MSPs.  The
average number of individual integrations included in each data point
varied with observational frequency but was typically 10 at 436 MHz,
and 25 or more at the higher frequencies.   The last two columns of
Table 1 give the spectral index and error for each pulsar. Figure 1
shows typical spectral plots and Figure 2 shows comparative
histograms of the spectral index distributions for the slow and
millisecond pulsars detected in the 70 cm survey.

For PSR B1620--26 (PSR J1623--2631) the 660 MHz flux density is
similar to that of FFB,  but the 1400 MHz flux density is higher,
giving a flatter spectrum (spectral index --1.5 compared to --2.5
). For this and other pulsars in the FFB study, their steeper spectral
indices result mainly from the inclusion of data points near 100
MHz. We also note that our total integration time was similar to that
used by FFB, and so we consider our value of similar statistical
significance.

\section{Discussion}

Interstellar scintillation results in variations of the pulse
intensity as a function of time and frequency.  Diffractive
interstellar scintillation (DISS) has a time scale, $T_{\rm{DISS}}$,
that is dependent upon the distance to the pulsar, the observing
frequency and the pulsar's transverse velocity (e.g., \cite{sc90a}).
The time scale and bandwidth of these intensity modulations increase as
the observing frequency increases. For MSPs at the lower frequencies,
the typical $T_{\rm{DISS}}$ was $\sim 10$ min and the typical
diffractive bandwidth was less than 1 MHz. By using a bandwidth of 32
MHz and integration times of 24 min at these low frequencies, we have
averaged over many scintils to obtain our flux density
measurements. The time scale for refractive interstellar scintillation
(RISS), $T_{\rm{RISS}}$, is typically of the order of a few days. We have
made sufficient observations to have averaged out the effects of RISS
in almost all of the pulsars.  The effects of DISS are more evident in
the higher moments of the flux density distributions. Typically, the
flux density distribution shows some positive skewness and kurtosis,
consistent with the exponential probability density function expected
to arise from scintillation. An effort was made to include all
observations of the highly scintillating pulsars to avoid the
observational bias that arises by observing pulsars only when
they are at a scintillation maximum.

Until the recent publication of MSP spectra by Kramer \etal (1998) 
\nocite{kxl+98} it was commonly accepted that the mean spectral index of 
MSPs was steeper than that of slow pulsars. The evidence for this had  
largely been based on flux density data presented in discovery papers.  
For example, the MSP spectral indices quoted by Lorimer \etal have a mean
of --2.0 $\pm$ 0.2 (\cite{lylg95} and references therein). 
The spectral index distribution for the 19 MSPs considered here has a 
mean spectral index is --1.9 $\pm$ 0.1, while the mean value for the slow 
pulsars in our sample of --1.72$\pm$0.04 is slightly steeper than the --1.6 
obtained by Lorimer et al. The difference between these two values can be  
understood in terms of selection effects. Lorimer's sample included all pulsars,  
including those found in high-frequency surveys, whereas the Parkes sample 
included only those detected at 436 MHz. Flat-spectrum pulsars are much easier to
detect at high frequencies because of reduced dispersion-measure
smearing, scattering and lower sky background temperatures.  Johnston
\etal (1992) \nocite{jlm+92} found, for example, that the median
spectral index for their 1500 MHz survey at Parkes was only --1.0. 

Our results agree well with Kramer et al.'s comparison of the spectra of 
the slow pulsars and many of the MSPs detected in the Parkes 70 cm survey. 
They determined the mean spectral indices to be --1.8$\pm$0.2 (MSPs) \
and --1.7$\pm$0.1 (slow pulsars). Although the individual spectra of MSPs 
published by Kramer \etal and those presented here show significant differences, 
the mean MSP spectral indices are within errors, while the mean spectra of 
slow pulsars are in excellent agreement.      

By combining flux density measurements taken at frequencies above 1400
MHz with published data, Malofeev \etal (1994 and references therein)
\nocite{mgj+94} were able to derive accurate spectral indices for 45 slow 
pulsars. If we consider only the 20 of  these pulsars that show no
evidence for a break in their spectra then the mean spectral  index
for these pulsars is --1.9. Kijak \etal (1998) \nocite{kkw+98} have
done similar work with flux  densities at frequencies between 1.41 and
4.85 GHz, deriving a mean spectral index of --1.9 for a sample of
144 slow pulsars. PSR J0437--4715, with a flux density of 11.6$\pm$0.1
mJy at 4.8 GHz was the only MSP for which we had flux density data
above 1600 MHz. With a spectral index of --2.1 between 1.4 and 4.8
GHz compared to --1.1 at lower frequencies, PSR J0437--4715 follows the
same trend as the slow pulsars in Kijak et al.'s sample.

The mean spectral index of the MSPs in our sample is $\sim$10\% steeper 
than that of the slow pulsars. To compare the distributions in more detail 
we did a Kolmogorov-Smirnov test that showed that there is a
28\% probability that the two samples are drawn from the same
distribution.  Therefore, the evidence that millisecond pulsars are
intrinsically steeper-spectrum objects is fairly weak. Similarly,
there is little evidence  when one compares the distributions
directly, as in Figure 2. We therefore suggest, in agreement with Kramer et al.'s
study, that the widely held view that MSP spectra are much steeper is unfounded, 
and due, in part, to the early methods for finding MSPs that {\it required} 
candidates to have steep spectra.

\section{A Simulated MSP High-frequency Survey}

Most of the more than 700 pulsars known today were discovered
in searches conducted at frequencies near 400 MHz, where pulsars are
relatively bright. The last decade has seen concerted efforts to
search for pulsars at higher frequencies where sensitivity is better
because  dispersion-measure smearing and scattering in the
interstellar medium are reduced.  One such survey is currently
under way at the Parkes observatory using the  13 beam H{\small I}
multibeam receiver system (\cite{cam96}). The survey plans to cover
the Galactic plane at 1400 MHz ($-5^{\circ} \leq b \leq +5^{\circ}$
and  $260^{\circ} \leq l \leq 20^{\circ}$). In the light of the new
spectral results presented in this paper we have modelled the pulsar
population and investigated the possible results of similar high-frequency 
surveys.  The primary aim of our simulations was to study the spatial
distribution of detectable MSPs in high-frequency surveys.     
     
The pulsar population synthesis modelling software we used was
developed by Lorimer \etal (1993) \nocite{lbdh93}. The model generates
an  evolved population of $10^5$ pulsars and conducts a search for
these  pulsars after taking into account selection effects. Our
selection of initial  galactocentric radii and heights above the
Galactic plane for each pulsar were  drawn from Gaussian probability
distributions with a scale length of 4.8 kpc for the disk, and a
z-height of 500 pc. The effects of the interstellar medium (ISM) on the
pulsar signals were  simulated using the Cordes \etal
(1991)\nocite{cwf+91} electron distribution  model. The pulse
broadening caused by scattering by the ISM, scattering time
$\tau_{scatt}$, was estimated by a power-law fit to the
DM-$\tau_{scatt}$  correlation (\cite{bwhv92}) and a simple
$\nu^{-4.4}$ scaling. The periods of MSPs were
drawn from a Gaussian distribution about a  mean value of log $P$
corresponding to 3.1 ms and a standard deviation in the log of 0.3.
Lorimer \etal (1993) \nocite{lbdh93} used a power-law expression
of the form  $L=\gamma P^{\alpha}{\dot{P}}^{\beta}$ to derive each
pulsar's luminosity, $L$, where $P$ is the pulsar's  period in
seconds, ${\dot{P}}$ its period derivative, and $\gamma$ is a scale
constant (\cite{ec89}).  The spread in luminosity about the luminosity
law was modelled by convolving this distribution with a Gaussian  in
the log. We found that the pulsar population synthesized in this way
was very sensitive to the form of the  luminosity law. Instead, we used
a simplified luminosity law in which we set $L=\gamma$, i.e. independent of 
$P$ and ${\dot{P}}$.

In order to model a realistic population we scaled the constant
$\gamma$ and the standard deviation of log $L$, $\sigma_{log L}$,
until the detailed model of the Parkes 70 cm  survey detected 19
MSPs. The initial spectral-index distribution was varied until the
observed MSP spectral-index distribution matched that presented in
this paper.  The mean spectral index of the parent population was
--1.95, with a standard deviation  of 0.50, and $\gamma$ was set to
$1.55$ mJy kpc$^{2}$ with $\sigma_{log L}$ set to 0.65.

We conducted a simulated 1400 MHz {\it all-sky} search of the evolved
population of MSPs. The search involved 35 min pointings of the
13-beam receiver assumed to  have an effective receiver temperature of
30 K. The back-end recording system had  $2\times 288$ MHz of
bandwidth, 3 MHz channels and a sampling rate of 4 kHz as described
in Camilo (1996) \nocite{cam96}. To investigate the likely spatial
distribution, our first model searched the entire sky. This model
search resulted in the  discovery of 172 MSPs with a minimum  flux
density of $\sim$0.7 mJy. This result suggests that such a survey
would result in a three-fold  increase in MSPs known. Figure 3 shows
that the distribution of the detected MSPs is far more  concentrated
on the plane than previous surveys. In fact 50\% of the detected MSPs
are within 11\arcdeg of the Galactic plane. This is easily
understood in terms of the higher space density of MSPs in that
region due to geometrical effects, and the relative insensitivity to 
scattering and high sky background temperatures of high-frequency surveys.

We modelled the current multibeam survey by restricting the search to
within $\pm 5^{\circ}$ of the plane and $260^{\circ} \leq l \leq
20^{\circ}$.  Our model survey detected 26 MSPs, 24 of which were new
discoveries.  Uncertainties in the MSP luminosity function and total
Galactic population, not to mention our simplified model of the Galaxy and 
multibeam survey system mean that the above estimate could be uncertain by  
a factor of two. Figure 4 shows the distribution of MSP dispersion
measures extends to a DM value of  $\sim$310 pc cm$^{-3}$. To date,
the only known MSPs with such high DMs were found in globular
clusters. The high DM tail shows that a high-frequency survey will
probe much deeper into the Galactic disk.

\section{Conclusion}

We have obtained reliable multi-frequency flux density
measurements for the 19 southern MSPs discovered in the Parkes 70 cm
survey, enabling us to determine the spectral index distribution for
this sample. We have compared this spectral-index distribution with
that of slow pulsars detected in the same survey. The spectral-index
of the millisecond pulsars was found to have a mean of $\alpha \sim
-1.9$, only slightly steeper than the mean $\alpha \sim -1.7$ for the slow
pulsars. A Kolmogorov-Smirnov test suggests only a 72\% probability
of the distributions differing.  This result adds to the growing evidence 
that MSPs have spectral properties similar to slow pulsars.  Our
simulation of high-frequency surveys similar to the Parkes multibeam survey, 
using the data presented in this paper, suggests that they will discover a
large number of MSPs. High-frequency surveys are therefore likely to have 
a significant effect on the known population of millisecond pulsars in 
the near future. 

\newpage
\begin{deluxetable}{cccccc}
\tablecolumns{6}
\tablewidth{0pc}
\tablecaption{Flux Densities and Spectral Indices for Southern MSPs}
\tablehead{
\colhead{PSR J}&
\colhead{$S_{436}$}& 
\colhead{$S_{660}$}& 
\colhead{$S_{1400}$}&
\colhead{$S_{1660}$}&
\colhead{$\alpha$}\\ 
\colhead{}&
\colhead{(mJy)}& 
\colhead{(mJy)}& 
\colhead{(mJy)}& 
\colhead{(mJy)}& 
\colhead{}   \\ }
\startdata
0034--0534
& 17(5)
& 5.5(5)
& 0.61(9)
& 0.56(9)
& --2.6(10)
\nl
0437--4715
& 550(100)
& 300(20)
& 137(3)
& 115(3)
& --1.1(5)
\nl
0613--0200
& \nodata
& 7.3(4) 
& 2.2(1) 
& 2.0(1) 
& --1.5(5)
\nl
0711--6830
&\nodata 
& 11(2) 
& 3.4(5) 
& 2.1(3) 
& --1.7(14)
\nl
1024--0719
&\nodata 
& 4.2(6) 
& 0.9(1) 
& 0.88(7) 
& --1.7(12)
\nl
1045--4509
& 15(3) 
& 8.5(3) 
& 1.9(1) 
&\nodata 
& --2.0(5)
\nl
1455--3330
& 9(1)
& 8(1)
& 1.2(1)
&\nodata 
& --1.8(11)
\nl
1603--7202
& 21(2)
& 17(1)
& 2.9(2)
& 2.3(2)
& --1.8(5)
\nl
1623--2631 
&\nodata  
& 7.2(2)		
& 2.0(3)
&\nodata 
& --1.5(2)
\nl
1643--1224
&\nodata 
& 16.0(5) 
& 3.3(1) 
& 3.0(1)
& --1.9(3)
\nl
1730--2304
&\nodata 
& 14(1) 
& 3.0(4) 
& 2.5(3) 
& --1.9(10)
\nl
1744--1134
& 18(2)
& 16(2)
& 2.0(2)
& 1.7(3)
& --1.8(7)
\nl
1804--2718
& 15(4)
&  8(1)
&  0.7(1)
&\nodata
&  --2.9(15)
\nl
1823--3021A
& 16(5)
&  6.8(7)
&  0.72(2)
&\nodata 
& --2.7(9)
\nl
1911--1114
& \nodata
& 7.1(5) 
& 0.7(1) 
& 0.63(4) 
& --2.6(7)
\nl
2051--0827
&\nodata
& 5.5(5) 
& 1.4(1) 
& 1.5(1) 
& --1.6(9)
\nl
2124--3358
& 17(4)
& 7.8(7) 
& 2.6(2) 
& 1.9(2)
& --1.5(8) 
\nl
2129--5718
& 14(2) 
& 8.4(7)
& 1.2(1)
& \nodata 
& --2.2(8)
\nl
2145--0750
& 100(30) 
& 36(4) 
& 7.0(9) 
& 5.4(8)
& --2.1(11)
\nl
\enddata
\end{deluxetable}

\newpage
\section*{Figure Captions}
\noindent
Figure 1. Spectra for nineteen millisecond pulsars illustrating that
MSP spectra between 400 and 1700 MHz are often well described by a
power law.

\noindent
Figure 2. Histograms of the spectral index distributions for 216 slow (P
$>$ 20 ms) pulsars and millisecond pulsars. The mean spectral index
for the slow pulsar distribution is $\alpha = -1.7$, whereas that for
the MSPs is $\alpha = -1.9$.

\noindent
Figure 3. Aitoff-Hammer projection of the distribution of MSPs
detected in a simulated high-frequency all-sky survey. A search
concentrated near the Galactic plane ($|b| \leq 10^{\circ}$) would
detect 44\% of the MSPs.

\noindent
Figure 4. Distribution of MSP dispersion measures for pulsars detected
in a simulated high-frequency survey ($260^{\circ} \leq l \leq 20^{\circ}$
and $|b| \leq 5^{\circ}$).

\noindent


\newpage

\begin{thebibliography}{}

\bibitem[Baars \etal  1977]{bgpw77}
Baars, J. W.~M., Genzel, R., Pauliny-Toth, I. I.~K., \& Witzel, A. 1977, {
  \aap}, {\rm 61}, 99.

\bibitem[Backer \etal  1982]{bkh+82}
Backer, D.~C., Kulkarni, S.~R., Heiles, C., Davis, M.~M., \& Goss, W.~M. 1982,
  { Nature}, {\rm 300}, 615.

\bibitem[Bhattacharya \etal  1992]{bwhv92}
Bhattacharya, D., Wijers, R. A. M.~J., Hartman, J.~W., \& Verbunt, F. 1992, {
  \aap}, {\rm 254}, 198.

\bibitem[Biggs \& Lyne 1992]{bl92}
Biggs, J.~D. \& Lyne, A.~G. 1992, { \mnras}, {\rm 254}, 257.

\bibitem[Camilo 1996]{cam96}
Camilo, F. 1996, in { High Sensitivity Radio Astronomy}, ed.\ N. Jackson,
  Cambridge University Press, In Press.

\bibitem[Camilo, Nice, \& Taylor 1993]{cnt93}
Camilo, F., Nice, D.~J., \& Taylor, J.~H. 1993, { \apjlett}, {\rm 412}, L37.

\bibitem[Clifton \& Lyne 1986]{cl86}
Clifton, T.~R. \& Lyne, A.~G. 1986, { Nature}, {\rm 320}, 43.

\bibitem[Clifton \etal  1992]{clj+92}
Clifton, T.~R., Lyne, A.~G., Jones, A.~W., McKenna, J., \& Ashworth, M. 1992, {
  \mnras}, {\rm 254}, 177.

\bibitem[Cordes \etal  1991]{cwf+91}
Cordes, J.~M., Weisberg, J.~M., Frail, D.~A., Spangler, S.~R., \& Ryan, M.
  1991, { Nature}, {\rm 354}, 121.

\bibitem[Emmering \& Chevalier 1989]{ec89}
Emmering, R.~T. \& Chevalier, R.~A. 1989, { \apj}, {\rm 345}, 931.

\bibitem[Erickson \etal  1987]{embh87}
Erickson, W.~C., Mahoney, M.~J., Becker, R.~H., \& Helfand, D.~J. 1987, {
  \apj}, {\rm 314}, {L}45.

\bibitem[Foster, Fairhead, \& Backer 1991]{ffb91}
Foster, R.~S., Fairhead, L., \& Backer, D.~C. 1991, { \apj}, {\rm 378}, 687.

\bibitem[Foster \etal  1995]{fcwa95}
Foster, R.~S., Cadwell, B.~J., Wolszczan, A., \& Anderson, S.~B. 1995, { \apj},
  {\rm 454}, 826.

\bibitem[Fruchter 1989]{fru89}
Fruchter, A.~S. 1989.
\newblock PhD thesis, Princeton University.

\bibitem[Hamilton, Helfand, \& Becker 1985]{hhb85}
Hamilton, T.~T., Helfand, D.~J., \& Becker, R.~H. 1985, { \aj}, {\rm 90}, 606.

\bibitem[Johnston \etal  1992]{jlm+92}
Johnston, S., Lyne, A.~G., Manchester, R.~N., Kniffen, D.~A., D'Amico, N., Lim,
  J., \& Ashworth, M. 1992, { \mnras}, {\rm 255}, 401.

\bibitem[Kijak \etal  1998]{kkw+98}
Kijak, J., Kramer, M., Wielebinski, R., \& Jessner, A. 1998, { \aap}, {\rm
  127}, 153.

\bibitem[Kramer \etal  1998]{kxl+98}
Kramer, M., Xilouris, K.~M., Lorimer, D.~R., Doroshenko, O., Jessner, A.,
  Wielebinski, R., Wolszczan, A., \& Camilo, F. 1998, { \apj}, {\rm }
\newblock In press.

\bibitem[Lorimer \etal  1993]{lbdh93}
Lorimer, D.~R., Bailes, M., Dewey, R.~J., \& Harrison, P.~A. 1993, { \mnras},
  {\rm 263}, 403.

\bibitem[Lorimer \etal  1995]{lylg95}
Lorimer, D.~R., Yates, J.~A., Lyne, A.~G., \& Gould, D.~M. 1995, { \mnras},
  {\rm 273}, 411.

\bibitem[Lyne \etal  1987]{lbm+87}
Lyne, A.~G., Brinklow, A., Middleditch, J., Kulkarni, S.~R., Backer, D.~C., \&
  Clifton, T.~R. 1987, { Nature}, {\rm 328}, 399.

\bibitem[Lyne \etal  1988]{lbb+88}
Lyne, A.~G., Biggs, J.~D., Brinklow, A., Ashworth, M., \& McKenna, J. 1988, {
  Nature}, {\rm 332}, 45.

\bibitem[Lyne \etal  1998]{lml+98}
Lyne, A.~G., Manchester, R.~N., Lorimer, D.~R., Bailes, M., D'Amico, N.,
  Tauris, T.~M., \& Johnston, S. 1998, { \mnras}, {\rm }
\newblock In press.

\bibitem[Malofeev \etal  1994]{mgj+94}
Malofeev, V.~M., Gil, J.~A., Jessner, A., Malov, I.~F., Seiradakis, J.~H.,
  Sieber, W., \& Wielebinski, R. 1994, { \aap}, {\rm 285}, 201.

\bibitem[Manchester \etal  1996]{mld+96}
Manchester, R.~N. \etal  1996, { \mnras}, {\rm 279}, 1235.

\bibitem[Navarro 1994]{nav94}
Navarro, J. 1994.
\newblock PhD thesis, California Institute of Technology.

\bibitem[Segelstein \etal  1986]{srs+86}
Segelstein, D.~J., Rawley, L.~A., Stinebring, D.~R., Fruchter, A.~S., \&
  Taylor, J.~H. 1986, { Nature}, {\rm 322}, 714.

\bibitem[Stinebring \& Condon 1990]{sc90a}
Stinebring, D.~R. \& Condon, J.~J. 1990, { \apj}, {\rm 352}, 207.

\end{thebibliography}
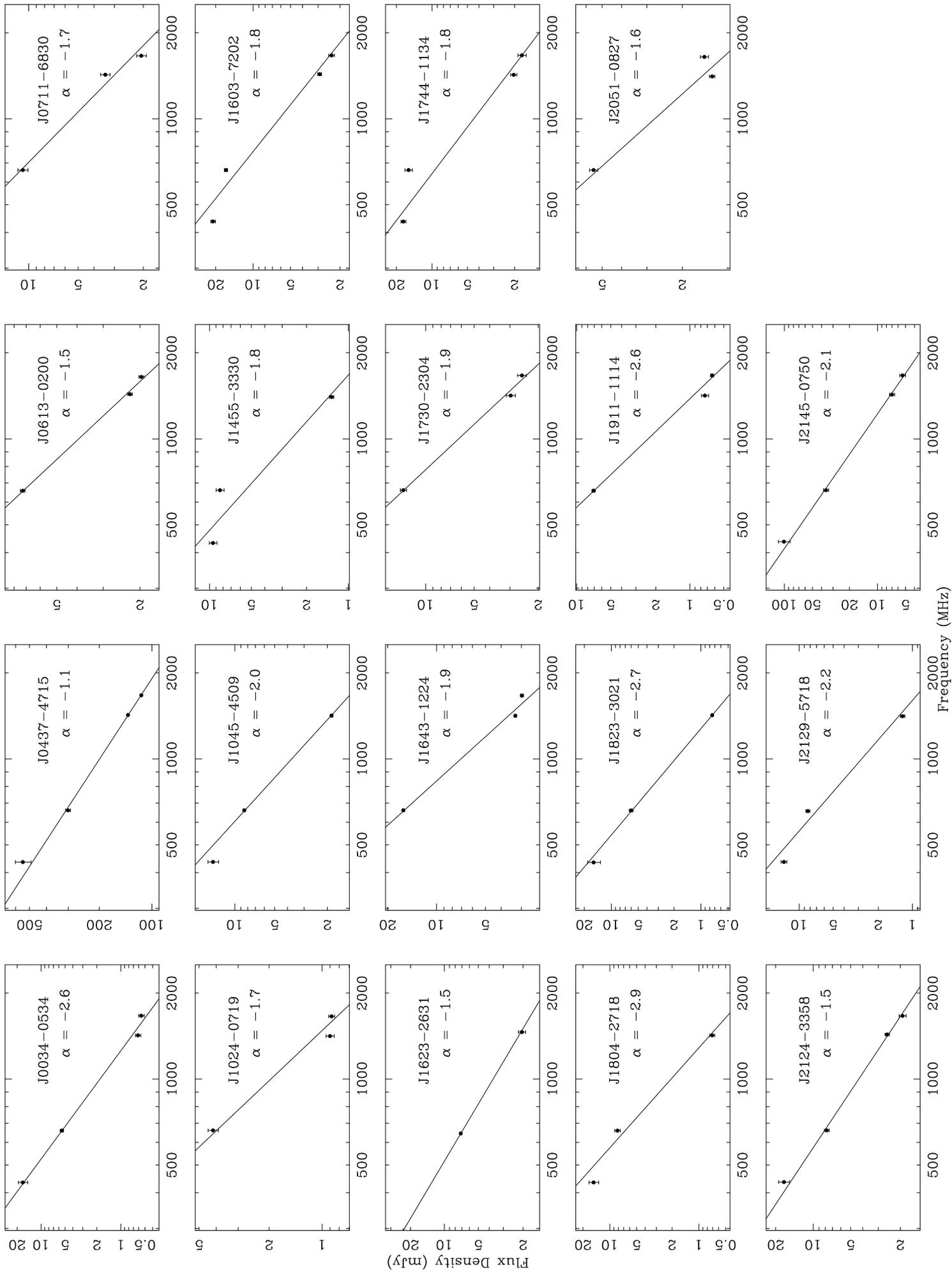
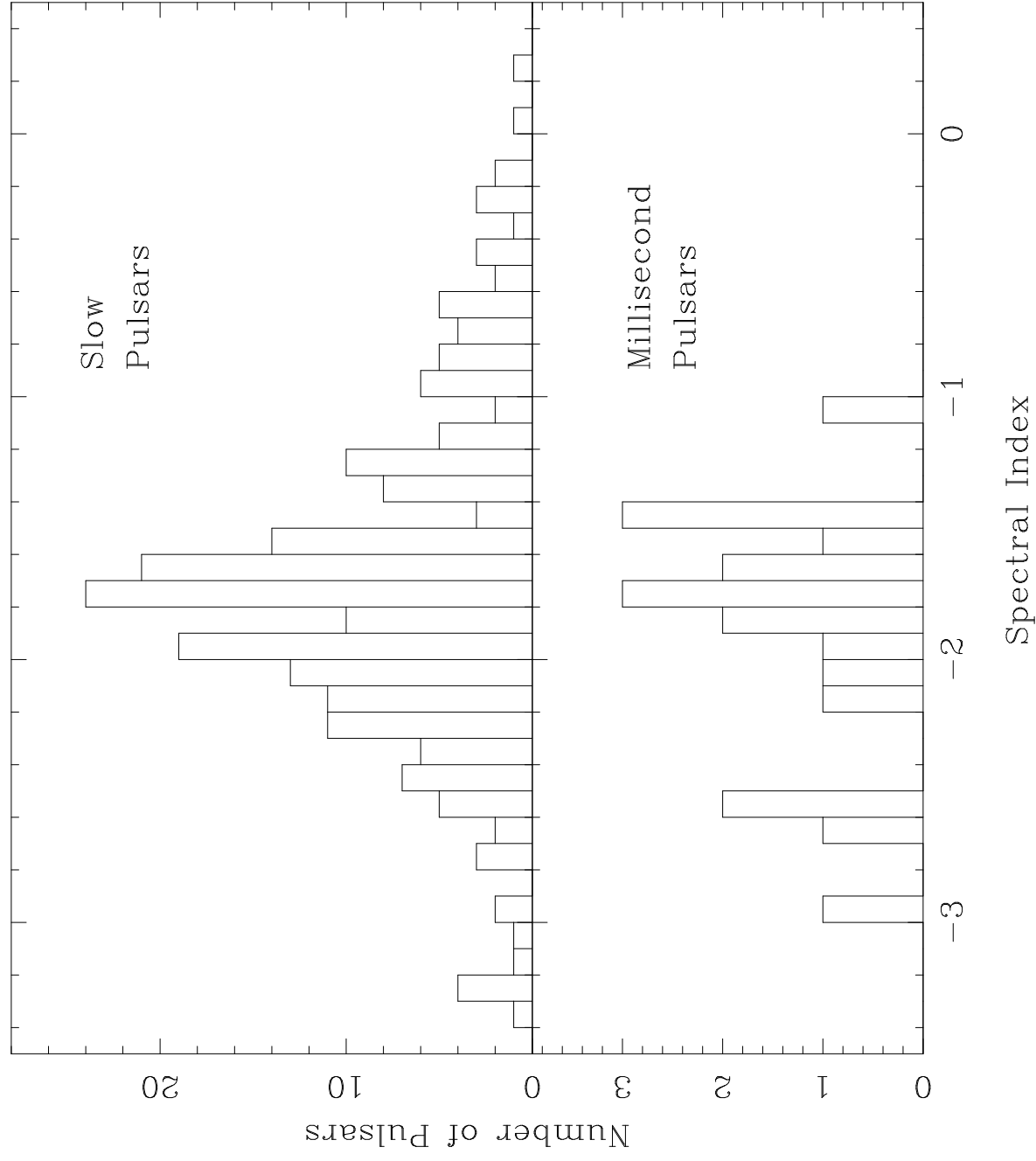
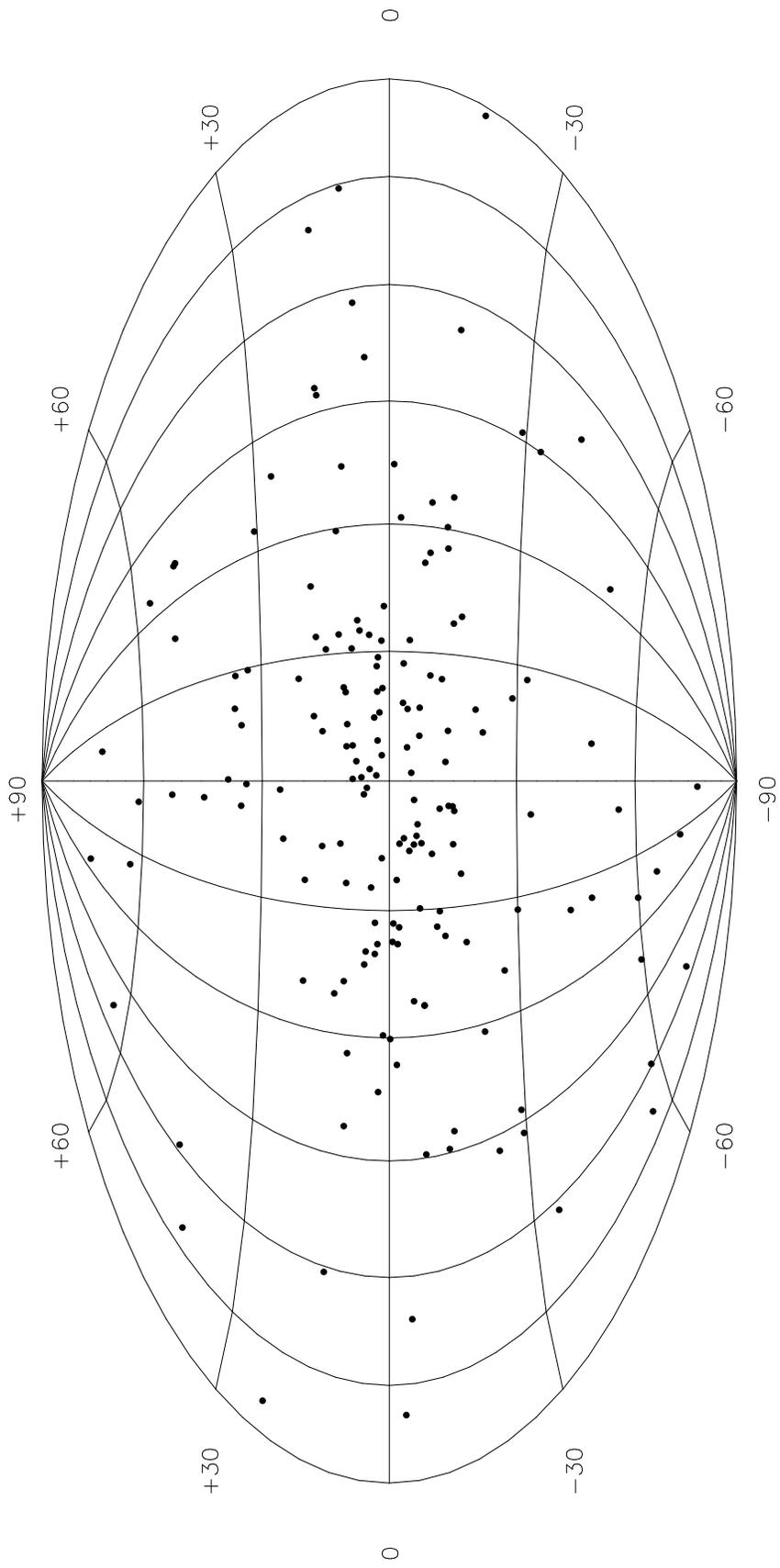
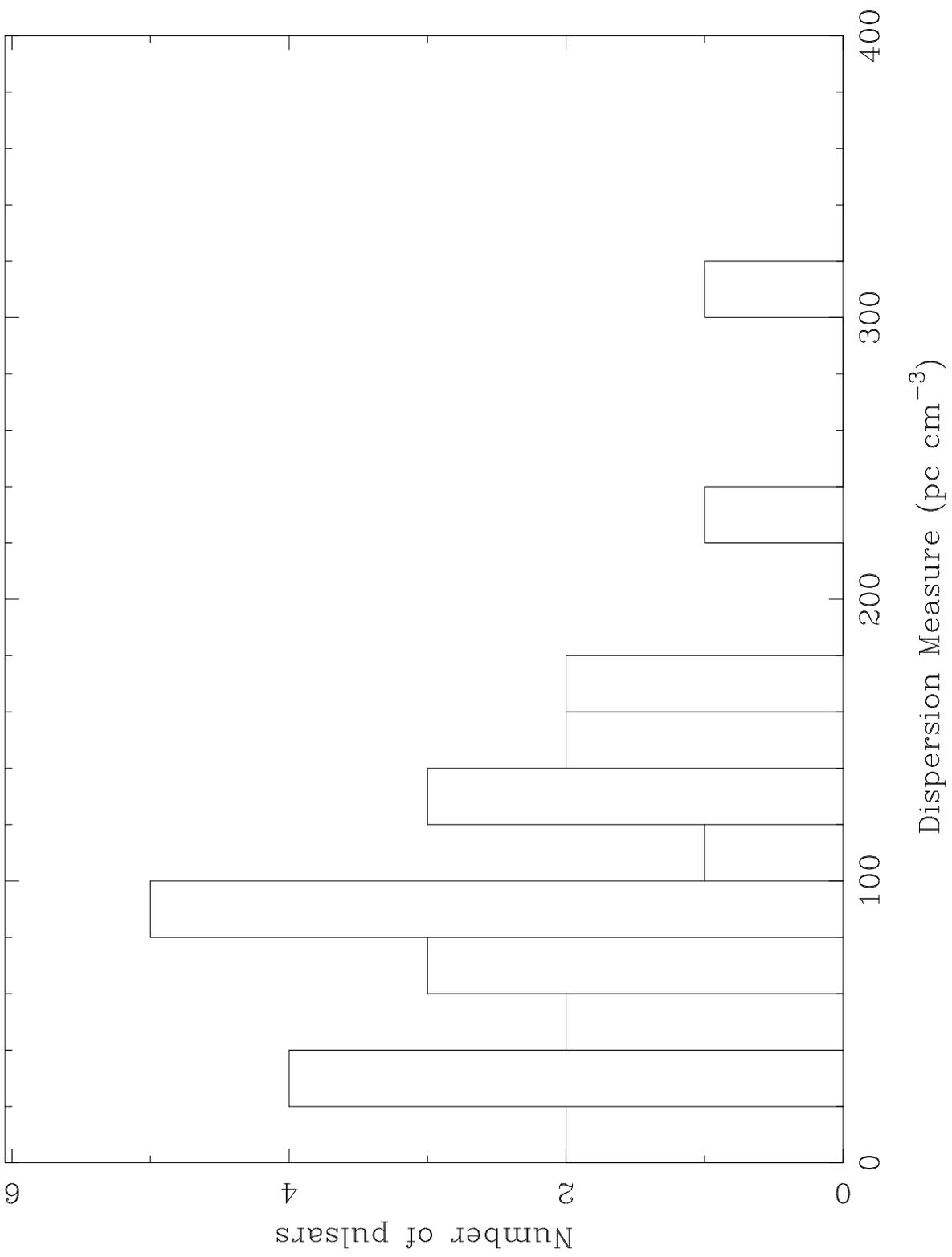
\end{document}